\documentclass[fleqn,twoside]{article}
\usepackage{espcrc2}

\usepackage{graphicx}
\usepackage{epsfig}
\usepackage[figuresright]{rotating}

\newcommand{\be}{\begin{equation}}
\newcommand{\ee}{\end{equation}} 
\newcommand{\ber}{\begin{eqnarray}}
\newcommand{\eer}{\end{eqnarray}}
 
\def\ket#1{\left| #1\right\rangle}
\newcommand{\AmS}{{\protect\the\textfont2
  A\kern-.1667em\lower.5ex\hbox{M}\kern-.125emS}}

\title{Light Mesons from Heavy B and Hyperon Decays}

\author{Patrick J. O'Donnell\address{Department of Physics, \\ 
        University of Toronto,\\ Toronto, Ont. Canada. M5S 1A7}%
        \thanks{Research supported by NSERC, grant number A3828. This
        talk is based on the papers with Alakabha Datta and Harry
        J. Lipkin:  - Phys. Lett. \textbf{B529} (2002)
        93-98 [hep-ph/0111336], Phys.Lett.\textbf{B540} (2002) 97-103 [hep-ph/0202235] and
       Phys.Lett. \textbf{B544} (2002) 145-153 [hep-ph/0206155]. These papers
        contain details and references that could not be
        included here.}
       }
       
\begin{document}

\begin{abstract}
Decays of heavy mesons and of heavy hyperons are used to
provide tests of the standard model and information about new mixing
schemes for the $\eta$ and $\eta^{'}$ mesons. These include the two
body decays $B_s \to J / \psi M$ and $B_d \to J / \psi M$, $B \to \eta
(\eta ^{'})K(K^\ast )$ and $\Lambda _b \to \Lambda \eta (\eta ^{'})$, 
semileptonic $D$ decays, and properties of radially excited mesons.

\vspace{1pc}
\end{abstract}

\maketitle
\section{Introduction}
In this talk, I consider the two body decays of heavy mesons
and of heavy hyperons $B_s \to J / \psi M$ \textit{and } $B_d \to J /
\psi M$, $B \to \eta (\eta ^{'})K(K^\ast )$, $\Lambda _b \to \Lambda
\eta (\eta ^{'})$ and the semileptonic $D$ decays. Tests of
the standard $\eta -\eta^{'}$ mixing and of properties of
radially excited mesons are given.
\section{Relations for two body $B \to J/\psi$ Decays and tests for $\eta 
-\eta^{'}$ mixing. }
Nonleptonic two--body decays $B^{0} \to J/\psi M$
and $B_{s} \to J/ \psi M$ are of particular interest since one of
these is the ``golden
channel" - $J$/$\psi K_{S}$, important for CP violation. In the
inactive spectator approach the analysis of large groups of different decays related by
symmetries is simplified since the bound $c \bar c$ pair is a singlet under color, isospin and
flavor SU(3) and is an eigenstate of $C$ and $P$. The two-meson
final state has an unique color coupling so that there are \textbf{selection rules} for all decays
without the spectator quark in the final state:
\begin{displaymath}
 A[B^{0} \to J/\psi M (\bar q s)] = 0,  
 A[B_{s} \to J/\psi M (\bar q d)] = 0
\end{displaymath}
The following relations for amplitudes are tests of these selection
rules:
\ber
A_{L}(B^{0} \to J/\psi \rho ^{o})&=&A_{L}(B^{0} \to J/\psi 
\omega ),\nonumber \\
A_{L}(B_{s} \to J/\psi \rho ^{o})&=&A_{L}(B_{s} \to J/\psi 
\omega )\nonumber \\
&=&A_{L}(B^{0} \to J/\psi \phi )=0   \nonumber
\eer
($L$ denotes any partial wave for the vector-vector final state)

All other decays are described by the two transitions: 
\ber
B(\bar b q)& \to J/\psi \bar d q   
&\to  J/\psi M(\bar d q)\nonumber \\
B(\bar b q) &\to J/\psi \bar s q &\to J/\psi M(\bar s q)\nonumber
\eer
where the
decay amplitudes are described as the product of a $\bar b$ decay
amplitude and a hadronization function for the combination of
a quark-antiquark pair making the final meson. Charge
conjugation symmetry implies no $SU(3)$ breaking in the
final state interactions and 
$$  
A( {B_s} \to J/\psi  \bar K^{*0})_L = F_{CKM}^{L}\cdot A( {B_d} \to
J/\psi  K^{*0})_L 
$$
where $ F_{CKM}^L$ is a factor depending on the ratios of the $CKM$
matrix elements and the ratio of various weak interaction diagrams 
contributing to $B_d$ and $B_s$ decays.
For the (dominant) tree-- and penguin--diagram contributions with a
charmed quark loop, $F_{CKM}^L=V_{cd}/V_{cs}$.

Including SU(3) symmetry gives an additional set of predictions 
\ber 
A_L( {B_d}^0 \to J/\psi \rho^o) 
&=&A_L( {B_d}^0 \to J/\psi \omega)\nonumber \\ 
&=&A_L( {B_s} \to J/\psi  \bar K^{*0})/{\sqrt 2}\nonumber \\
A_L( {B}_s \to J/\psi \phi) 
&=& A_L( {B}^0 \to J/\psi  K^{*0})
\nonumber
\eer
\section{B decays into charmonium and a pseudoscalar meson}
Decays involving  $\eta $ or the $\eta ^{'}$ mesons in the final state
have been unexplained by the standard treatments of these
decays. Mixing, in general, involves 
four different radial
wave functions and cannot be described by diagonalizing a simple 
$2 \times 2$ matrix with a single mixing angle. 
Write the normalized $\eta-\eta^{\prime}$ wave functions as 
\ber 
\ket{\eta} & = & \cos{\phi}\ket{N} -\sin{\phi}\ket{S}\nonumber\\ 
\ket{\eta^{\prime}} & = & \sin{\phi^{\prime}}\ket{N^{\prime}} + \cos{\phi^{\prime}}\ket{S^{\prime}}\nonumber
\eer
where $\ket{N}$, $\ket{N^{\prime}}$, $\ket{S}$ and $\ket{S^{\prime}}$
are respectively arbitrary isoscalar non-strange and strange
quark-antiquark wave functions. When, traditionally, the
$\eta-\eta^{\prime}$ mixing is described by a single mixing angle, 
\begin{displaymath}
  \ket{N}  =  \ket{N^{\prime}},\ 
\ket{S} = \ket{S^{\prime}},\
\phi = \phi^{\prime},
\end{displaymath}
and, including phase space effects, gives the experimentally measurable quantities,
\ber
r_d & \equiv &\frac
{p_{\eta^{\prime}}^3\Gamma( {\bar B}^0 \to J/\psi \eta) }
{p_{\eta}^3\Gamma( {\bar B}^0 \to J/\psi \eta^{\prime})} = \cot^2\phi,\ \nonumber
\eer
\ber
r_s & \equiv &\frac
{p_{\eta^{\prime}}^3\Gamma( {\bar B_s}^0 \to J/\psi \eta) }
{p_{\eta}^3\Gamma( {\bar B_s}^0 \to J/\psi \eta^{\prime})} =
\tan^2\phi,\ \nonumber
\eer 
with
\ber 
r & = & \sqrt{r_dr_s}=1.\ \nonumber 
\eer 
\textit{\textbf{Any large deviation of $r$ from 1 would indicate evidence of non-standard $\eta -\eta^{'}$ mixing.} }

If the pseudoscalar mesons have the same radial wave functions
$SU(3)$ symmetry relates the amplitudes:
\ber
A(B_{d} \to J/\psi N)&=& A(B_{s} \to J/\psi\bar{K^{0}})/\sqrt(2)\nonumber \\
&=& A(B_{s} \to J/\psi \pi^{0}),\nonumber\\
A(B_{s} \to J/\psi S)&=&A(B_{d} \to J/\psi K^{0})\nonumber 
\eer
and gives sum rules for standard mixing that are independent of the mixing angle 
\ber
|A(B_d \to J/\psi \eta)|^2 +
|A(B_d \to J/\psi \eta^{\prime})|^2 \nonumber \\
=|A( {B_s} \to J/\psi  \bar K^0)|^2/2 \nonumber\\
|A(B_s \to J/\psi \eta)|^2 +
|A(B_s \to J/\psi \eta^{\prime})|^2\nonumber \\ 
= |A( {B_d} \to J/\psi  K^0)|^2.
\ \nonumber
\eer
Charge conjugation relates the ratios of
$B_d$ and $B_s$ decays to $J/\psi \eta$ and $J/\psi \eta^{\prime}$, 
\ber r_\eta &  = &\frac {p_{B_s\eta}^3\Gamma( {B_d} \to J/\psi \eta) }
{p_{B_d\eta}^3\Gamma({B_s} \to J/\psi \eta)} =
(F_{CKM})^2\cdot\cot^2\phi,\ 
\nonumber
\eer 
\ber r_\eta^{\prime}& = &\frac {p_{B_s\eta^{\prime}}^3\Gamma( {\bar B}^0 \to J/\psi \eta^{\prime}) } {p_{B_d\eta^{\prime}}^3\Gamma( {\bar B_s} \to J/\psi \eta^{\prime})} = 
( F_{CKM})^2\cdot \tan^2\phi,\ 
\nonumber
\eer
where $ F_{CKM}= A ( {\bar b \to J/\psi  \bar d)}/ A( {\bar b
    \to J/\psi \bar s)}$, and predicts
\ber
r_B & = & \sqrt{r_\eta r_\eta^{\prime}}= (F_{CKM})^2.\
\nonumber 
\eer 
We can  also describe the \textit{branching ratios for eight
transitions in terms of three parameters} $F_{CKM}$, $\phi$ and an
overall normalization. 

If these relations hold experimentally, the
standard mixing and the value of the mixing angle will be confirmed
and established, the validity of SU(3) symmetry for these transitions
will be confirmed, and the value of $F_{CKM}$ will determine the ratio
of the penguin to tree contributions to the decay $B_d \to J/\psi K_S$
which is the ``golden channel" for CP violation
experiments. Otherwise, we get clues to new physics.
\section{$\eta -\eta^{'}$ mixing in semileptonic charmed meson decays}
In addition to $B$ decays we can also use $D(D_{s}) \to \eta (\eta
^{'})l\nu $ to get clean tests for mixing. Define the two ratios 
\ber
r_{\eta} & = &\frac{\Gamma(D \to \eta l \nu)}{\Gamma(D_s \to \eta l \nu)} 
\nonumber\\
r_{\eta^{\prime}} & = &\frac{\Gamma(D \to \eta^{\prime} l \nu)} {\Gamma(D_s \to \eta^{\prime} 
l \nu)} \
\nonumber
\eer
In the U spin limit,
\ber
r_D & = & \sqrt{r_{\eta}r_{\eta^{\prime}}} =1\
\nonumber
\eer
A \textit{large deviation of $r_D$ from 1 would indicate
evidence of non standard $\eta-\eta^{\prime}$ mixing} since such a
deviation is unlikely to originate from U spin breaking.

Including $q^{2}$ phase space gives the two ratios 
\ber
R_d(q^2) & = &\frac{p_{\eta^{\prime}}^3}{p_{\eta^3}}
 \frac{\frac{d\Gamma}{dq^2}(D_d \to \eta l \nu)}
                    {\frac{d\Gamma}{dq^2}(D_d \to \eta^{\prime} l
                      \nu)}= \cot^2{\phi}, \nonumber\\
R_s(q^2) & = &\frac{p_{\eta^{\prime}}^3}{p_{\eta^3}}
 \frac{\frac{d\Gamma}{dq^2}(D_s \to \eta l \nu)}
                    {\frac{d\Gamma}{dq^2}(D_s \to \eta^{\prime} l
                      \nu)}= \tan^2{\phi},\ 
\nonumber
\eer 
with standard mixing, and  
\ber 
R & = & R_d(q^2)R_s(q^2) =1 \nonumber
\eer 
for any value of $q^2$. Again, \textit{a deviation of $R$ from 1 or a 
$q^2$ dependence for $R$, $R_d$ and $R_s$ would indicate evidence of non-standard mixing}. 
\section{Charmless B Decays to Final States with Radially Excited Vector 
Mesons }
A meson in the final state of a nonleptonic decay
could be in a radially excited state. Define the ratios 
\ber
R_{\rho^+} & = & BR({\bar{B}^0} \rightarrow \pi^- \rho^{+ \prime} )/
BR({\bar{B}^0} \rightarrow \pi^- \rho^{+} ),
\nonumber
\eer
\ber
R_{\rho^0} & = & BR({\bar{B}^-} \rightarrow \pi^- \rho^{0 \prime} )/
BR({\bar{B}^-} \rightarrow \pi^- \rho^{0} ),
\nonumber
\eer
\ber
R_{\omega} & = & BR({\bar{B}^-} \rightarrow \pi^- \omega' )/BR({\bar{B}^-} \rightarrow \pi^- 
\omega),
\nonumber
\eer
\ber
R_{\phi} & = & BR({\bar{B}^-} \rightarrow \pi^- \phi' )/BR({\bar{B}^-} \rightarrow \pi^- 
\phi),
\nonumber
\eer
where $ \rho^{\prime}, \omega^{\prime} $ and $\phi^{\prime}$
are radially excited states.
%%%%%%%%%%%%%%%%%%%%%%%%%Remove for space%%%%%%%%%%%%%%%%%%%%%%%%%%%
%Neglecting relative Fermi momentum of the b quark and the
%spectator quark in the $B$ meson, the quark transition for these
%processes is
%  \begin{displaymath}
%   b \rightarrow \pi^-(\vec p) u(-\vec p), 
%   b \rightarrow \pi^0(\vec p) s(-\vec p)
%  \end{displaymath}
%where the b quark is at rest and $\vec p$ denotes 
%the final momentum of the $\pi$. 
%%%%%%%%%%%%%%%%%%%%%%%%%%%%%%%%%%%%%%%%%%%%%%%%%%%%%%%%%%%%%%%%%%%%%%%
%%%%%%%%%%%%%%%%%%%%%%%Figure gone to make space%%%%%%%%%%%%%%%%%%%%%%%
%\begin{figure}[htb]
%   \centerline{\epsfysize 2.2 truein \epsfbox{fac.ps}}
%\centerline{\includegraphics[width=2.2in]{fac.ps}}
%   \caption{Factorization for the decay $B \to M \pi$. }
%\label{fac}
%   \end{figure}

%\begin{figure}[htbp]
%\centerline{\includegraphics[width=3.94in,height=1.75in]{talkbeach1.eps}}
%\label{fig1}
%\end{figure}
%%%%%%%%%%%%%%%%%%%%%%%%%%%%%%%%%%%%%%%%%%%%%%%%%%%%%%%%%%%%%%%%%%%%%%%%%%
We need to diagonalize the mass matrix,
\begin{eqnarray*}
  \lefteqn{< q_a'\bar{q}_b',n'|M|q_a \bar{q}_b,n> = } \\
& & (m_a +m_b +E_n) \\
& & +\delta_{aa'}\delta_{bb'}\frac{B}{m_am_b}
{ \vec{s}_a \cdot \vec{s}_b} \psi_n(0)\psi_{n'}(0),\
\end{eqnarray*}
where $\vec{s}_{a,b}$ and $m_{a,b}$ are the quark spin operators and masses.
Here $n=0,1,2$ and the basis states for the isovector  mesons are chosen as
$\ket{N,I=1,I_3=1}=-\ket{ u \bar{d}}$,
$\ket{N,I=1,I_3=0}=\ket{ u \bar{u} - d \bar{d}}/\sqrt{2}$ and
$\ket{N,I=1,I_3=-1}=\ket{ d \bar{u}}$. $E_n$ is the excitation energy of the $n^{th}$ 
radially excited state and $B$ is the strength of the hyperfine interaction. 
%%%%%%%%%%%%%%%%%%%%%%%%%%%%%%Remove for space%%%%%%%%%%%%%%%%%%%%%%%
%As an example, we show a mixing solution in an harmonic potential model:
%\begin{table}[thb]
%\caption{Eigenvalues and Eigenstates for the $\rho$ system-Harmonic potential }
%\begin{center}
%\begin{tabular}{|c|c|c|c|c|}
%\hline
%& $N_0$ & $N_1$ & $N_2$ \\
%\hline
%$\rho(0.768)$ &
%0.99 &0.12& -0.07 \\
%\hline
%$\rho(1.545)$ &
%0.11 &-0.97& 0.20 \\
%\hline
%$\rho(2.370)$ &
%0.09 &-0.19& 0.98 \\
%\hline
%\end{tabular}
%\end{center}
%\end{table}
For $\omega - \phi $ mixing, the mass matrix has the
additional flavor mixing term 
\ber 
& + & \delta_{ab} \delta_{a'b'}
\frac{A}{m_am_b} \psi_n(0)\psi_{n^{\prime}}(0).\ \nonumber 
\eer 
(The basis states for the isoscalar mesons $|N>_n=| u \bar{u} + d
\bar{d}>_n/\sqrt{2}$ and $|S>_n=|s \bar{s}>_n$ for the non--strange
and strange wave functions).
%%%%%%%%%%%%%%%%%%%%%%%%%%%Remove table for space%%%%%%%%%%%%%%%%%%%%%%%
%\begin{table}[thb]
%\caption{Eigenvalues and Eigenstates for the $\omega-\phi $ system- Harmonic potential }
%\begin{center}
%\begin{tabular}{|c|c|c|c|c|c|c|c|}
 %     \hline & $N_0$ & $N_1$ & $N_2$ & $ S_0$ & $S_1$& $ S_2$\\ \hline
  %         $\omega(.783)$ & .98 &.15& -.08 & -.03& .01& -.01 \\ \hline
   %           $\phi(1.05)$ & .03 &.03& -.01 & .99 &.09 &-.05 \\ \hline
    %     $\omega(1.57)$ & -.13 & .95 & .26& -.01& -.14 & .01 \\ \hline
     %       $\phi(1.68)$ & .03 &-.12& -.07 & .08 &-.98 &-.14 \\ \hline
%\end{tabular}
%\end{center}
%\end{table}
%%%%%%%%%%%%%%%%%%%%%%%%%%%%%%%%%%%%%%%%%%%%%%%%%%%%%%%%%%%%%%%%%%%%%%%%
Different models have only a slight effect.
Typically,
\begin{displaymath}
  R_{\rho^+}:R_{\rho^0}: R_{\omega}: R_{\phi} = 2:2:2.5:6
\end{displaymath}
%%%%%%%%%%%%%%%%%%%%%%%%Remove table for space%%%%%%%%%%%%%%%%%%%%%%%%%%%
%\begin{table}[thb]
%\caption{Ratios of branching ratios for different confining potentials}
%\begin{center}
%\begin{tabular}{|c|c|c|c|}
%\hline
%Ratio & Linear & Quadratic & Quartic  \\
%\hline
%$R_{\rho^+}$ &
%$2.3$ & $2.0$ & $1.9$ \\
%\hline
%$R_{\rho^0}$ &
%$2.3$ & $2.0$ & $1.9$ \\
%\hline
%$R_{\omega}$ &
%$3.5$ & $2.5$ & $1.7$ \\
%\hline
%$R_{\phi}$ &
%$6.7$ & $6.2$ & $5.2$ \\
%\hline
%\end{tabular}
%\end{center}
%\end{table}
%%%%%%%%%%%%%%%%%%%%%%%%%%%%%%%%%%%%%%%%%%%%%%%%%%%%%%%%%%%%%%%%%%%%%%%%%
There are two main conclusions:-- \textit{Although mixing
  between radially excited states and the ground state is small,
  decays to excited states are enhanced} and, in particular, 
\textit{there is a large enhancement for R}$_{\phi }$. (This decay is
suppressed in the standard model)
\section{Non Standard $\eta -\eta^{'}$ mixing and 
Nonleptonic $B$ and $\Lambda_b$ Decays}
Mixtures of the ground state and radially excited $q \bar q$ states
can alter the high momentum behavior of the $\eta $ and $\eta ^{'}$
wave functions.  The decays $B \to \eta(\eta^{\prime})K(K^*)$ are
dominated by the penguin diagrams since the tree term is color and CKM
suppressed. With factorization we can have the kaon leaving
the weak vertex with its full momentum and the remaining quark
combining with the spectator quark to form the final
$\eta(\eta^{\prime})$ meson, or the $\bar{s}$ quark in the QCD penguin
combining with the $s$ quark from the $ b \to s$ transition to form
the $\eta(\eta^{\prime})$. Another possibility is one in which a
$q\bar{q}$ pair (where $ q=u,d, s$) appearing in the same current in
the effective Hamiltonian, hadronizes to the $\eta(\eta^{\prime})$.
This is often referred to the OZI suppressed term. However, it is not as simple
as this when the $\eta^{\prime}$ is in the final state because the OZI
suppressed terms add constructively.  We expect the OZI suppressed
terms to be more important in $B \to K P$ than in $B \to K V$ decays
since in $J/\psi$ and $\Upsilon$ decays we know that the OZI-forbidden
process requires three gluons for coupling to a vector meson and two
gluons for coupling to a pseudoscalar. Thus the
OZI suppressed terms should have a smaller contribution to the $B \to K
\rho^0(\omega)$ and $B \to K \phi$ decays than to $ B \to K \eta$ and
$B \to K \eta^{\prime}$ decays.
%%%%%%%%%%%%%%%%%%%%%%%%%%%Remove figure for space%%%%%%%%%%%%%%%%%%%
%\begin{figure}[htb]
%   \centerline{\epsfysize 2.2 truein \epsfbox{eta.ps}}
%\centerline{\includegraphics[width=2.9in]{eta.ps}}
% \caption{The  diagrams contributing to $B \to
% \eta(\eta^{\prime})K(K^*)$ 
% decays. The dashed line represents a gluon or a $\gamma$
%     or a Z boson. Tree diagrams are not shown. For the decays
%     $\Lambda_b \rightarrow \Lambda \eta(\eta^{\prime})$ diagram (a)
%     is absent.}
%   \end{figure}
%%%%%%%%%%%%%%%%%%%%%%%%%%%%%%%%%%%%%%%%%%%%%%%%%%%%%%%%%%%%%%%%%%%%%%
In the absence of a model independent extraction of $\phi$ we will use
the Isgur mixing, $\phi=45^0$, as our standard.
\ber
\ket{\eta}_{std} & = & \frac{1}{\sqrt{2}}\left[N_0 -S_0 \right],
\nonumber\\ 
\ket{\eta^{\prime}}_{std} & = & \frac{1}{\sqrt{2}}\left[N_0 +S_0
\right]. \ \nonumber 
\eer
For the $\eta-\eta^{\prime}$ system,
including radial excitations, we diagonalize the mass matrix as
in the $\omega - \phi$ case. The value of the parameter $A$ is
significantly changed. Nonet symmetry is broken; members of the
pseudoscalar nonet do not all
have the same radial wave function. This non--standard mixing gives
important differences for the
nonleptonic decays $B \to \eta (\eta ^{'})K(K^{\ast })$. 
%%%%%%%%%%%%%%%%%%%%%%%%%%%%Remove table for space%%%%%%%%%%%%%%%%%
%\begin{table}[thb]
%\caption{Eigenvalues and eigenstates for the $\eta-\eta^{\prime}$
%  system with $A=0.11m_u^2$ } 
%\begin{center} 
%\begin{tabular}{|c|c|c|c|c|c|c|c|}
%\hline
%Quartic  & $N_0$ & $N_1$ & $N_2$ & $ S_0$ & $S_1$ & $ S_2$\\ 
%\hline $\eta(.547)$ & .76 &-.29& .20 & -.44 & .25& -.20 \\ 
%\hline $\eta^{\prime}(.940)$ & .62 &.35& -.18 & .66 &-.13 &.09 \\
%\hline 
%\end{tabular}
%\end{center}
%\end{table}
%%%%%%%%%%%%%%%%%%%%%%%%%%%%%%%%%%%%%%%%%%%%%%%%%%%%%%%%%%%%%%%%%%%
\ber 
\ket{\eta}_g & = & 0.85\ket{\eta}_{std} +0.23
\ket{\eta^{\prime}}_{std},\nonumber \\ 
\ket{\eta^{\prime}}_g & = & 0.91\ket{\eta^{\prime}}_{std} -0.025
\ket{\eta}_{std}. \nonumber
\eer 
For $B \to \eta (\eta^{'}) K( K^{\ast})$ we find that they 
are \textit{dominated by the penguin diagrams} and that the
\textit{tree term is color and CKM suppressed.} The ratio $r$ can be a
small as $0.2$. 
In the absence of OZI terms there are definite predictions
about the branching ratios $B \to \eta K$/$B \to \eta ^{'}K$ and $B
\to \eta K^{\ast }$/$B \to \eta ^{'}K^{\ast }$. A parity selection
rule for the decays $B \to \eta (\eta ^{'})K^{(\ast ) }$ fixes the
relative phase between the amplitudes for the strange and
non--strange penguins causing interference.
\ber
R_{K} & \approx & \left|\frac{f_KF^{+}_{\eta}+ f_{\eta}^sF^{+}_{K}}
  {f_KF^{+}_{\eta^{\prime}}+ f_{\eta^{\prime}}^sF^{+}_{K}}\right|^2
\sim 0.03,\nonumber\\
R_{K^{*}} & \approx & \left|\frac{f_{K^{*}}F^{+}_{\eta}- f_{\eta}^sF^{+}_{K^{*}}} {f_{K^{*}}F^{+}_{\eta^{\prime}}- 
f_{\eta^{\prime}}^sF^{+}_{K^{*}}}\right|^2\sim \frac{1}{R_{K}}=33,\nonumber
\eer
the last from neglecting form factor differences. The small value for
$R_K$ is consistent with the
current experimental limits.

For $\Lambda_b$ only one diagram contributes:
\ber
R_{\Lambda} & \approx &
\left|\frac{f_{\eta}^sF_{\Lambda}} 
{f_{\eta^{\prime}}^sF_{\Lambda}}\right|^2\sim 1\,\nonumber
\eer
Additional sum rules connect $B$ decays to $K
\eta({\eta^{\prime}})$ and $ K \pi$ final states. One such sum rule is,
\ber
R & = & \frac{\Gamma[B^{\pm} \to K^{\pm} \eta^{\prime}]
+
\Gamma[B^{\pm} \to K^{\pm} \eta]}
{\Gamma[B^{\pm} \to K^{\pm} \pi^0]}  \le  3 \nonumber
\eer
Standard mixing gives $R \approx $ 3. Non--standard $\eta -\eta^{'}$ mixing gives 
$R \approx 6$, \textit{consistent with experiment}. Similarly, for $K^*$ final
states,
\ber 
\frac{\Gamma[B^{\pm} \to K^{*\pm} \eta]} {\Gamma[B^{\pm} \to
  K^{*\pm}
  \pi^0]}  & \approx &  |\sqrt{2} +1|^2 \approx 6 \nonumber,\\
\frac{\Gamma[B^{\pm} \to K^{*\pm} \eta^{\prime}]} {\Gamma[B^{\pm} \to
  K^{*\pm} \pi^0]} &\approx & |\sqrt{2} -1|^2 \approx
\frac{1}{6}.
\nonumber 
\eer 
New radial admixtures in the wave functions of the $\eta $ and $\eta
^{'}$ can increase the $B \to K\eta ^{'}$ decay but not the $B \to
K^{\ast }\eta ^{'}$ decay, as in
other OZI-violating models. Thus there is a \textit{suppression for $B
  \to K^{\ast} \eta^{'}$ decay}, an \textit{enhancement for $B \to
  K^{\ast }\eta $ decay relative to $B \to K^{\ast} \pi $} and an
\textit{enhancement for $B \to K \eta^{'}$ decay relative to $B \to K
  \pi $}.

For the OZI suppressed contribution, 
\ber 
\frac{A^{OZI}_{\eta^{\prime}}K(K^*)} {A^{OZI}_{\eta}K(K^*)} &
\approx & \frac{(1+\frac{1}{\sqrt{2}})} {(1-\frac{1}{\sqrt{2}})} \sim
\frac{A^{OZI}_{\eta^{\prime}}\Lambda} {A^{OZI}_{\eta} \Lambda} \sim
6\nonumber 
\eer 
A complete calculation gives 
%%%%%%%%%%%%%%%%%%%%%%Remove table for space%%%%%%%%%%%%%%%%%%%%%%%%
%\begin{table}[thb] 
%\caption{Branching ratios(BR) for $B \to \eta(\eta')K(K^*)$ decays and
%$\Lambda$ decays. Only CLEO results are shown; For the first item,
%BABAR reports a similar result.} 
%\begin{center} 
%\begin{tabular}{|c|c|c|} 
%\hline 
%Process & Exp. BR ($\times 10^{6})$ & Th. BR $(\times 10^{6})$ \\ 
%\hline 
%$B^-\rightarrow K^- \eta{\prime}$ & $(80 ^{+10}_{-9} \pm 7)$
% &$
%62$ \\ 
%\hline 
%$B^-\rightarrow K^- \eta$ & $ < 6.9 $ &$ 2.2$ \\
%\hline
%$B^-\rightarrow K^{-*} \eta^{\prime}$ & $ < 35$ &$ 1.6 $ \\
%\hline 
%$B^-\rightarrow K^{-*} \eta $ & $(26.4 ^{+9.6}_{-8.2} \pm 3.3)$ &$ 9 $ \\ 
%\hline 
%$\Lambda_b \rightarrow \Lambda \eta $ & - &$ 4.6$ \\ 
%\hline 
%$\Lambda_b \rightarrow \Lambda \eta^{\prime} $ & - &$ 12$ \\ 
%\hline 
%\end{tabular}
%\end{center}
%\end{table}
%%%%%%%%%%%%%%%%%%%%%%%%%%%%%%%%%%%%%%%%%%%%%%%%%%%%%%%%%%%%%%%%%%%%
\begin{displaymath}
  R_K  = 0.035, R_{K^*} =  5.6, R_{\Lambda}  =  0.4.
\end{displaymath}
in reasonable agreement with experiment.
\section{Summary and Conclusions }
Two body decays of $B_{d} \to J$/$\psi M$ and
$B_{s} \to J$/$\psi M$ have interesting relations among decay
amplitudes. We propose tests for the standard $\eta -\eta ^{'}$ mixing.
$SU(3)$ symmetry gives additional relations among the
eight CP eigenstates $J$/$\psi K_{S}$, $J$/$\psi \eta $, $J$/$\psi
\eta $' and $J$/$\psi \pi ^{0}$ and the value of the $\eta -\eta ^{'}$
mixing angle. Semileptonic $D$ decays can also give important
information about the standard $\eta -\eta ^{'}$ mixing. Weak decays
of a B meson to final S-wave radially excited give ratios of
nonleptonic decays $B \to \rho^{'} \pi/{B \to \rho \pi}, B \to
\omega^{'} \pi /{B \to \omega \pi}$ and $B \to \phi^{'} \pi /{B \to
  \phi \pi }$ transitions to the excited states comparable,
or enhanced relative, to ground state transitions. Radial mixing
in the $\eta - \eta ^{'}$ system leads to appreciable effects that gives a good
description of the existing data on $B \to \eta (\eta^{'})$. There are
predictions for as yet unobserved decays unique to our model.
Finally another unique prediction of our model is that there is a
modest enhancement of $\Lambda_b \to \Lambda \eta{'}$ relative to
$\Lambda_b \to \Lambda \eta$, unlike in the B system.
\end{document}